\documentclass{PoS}

\title{COLLIER -- \\ A fortran-library for one-loop integrals}
\ShortTitle{COLLIER}

\author{Ansgar Denner\\
        Universit\"at W\"urzburg\\
        E-mail: \email{denner@physik.uni-wuerzburg.de}}

\author{Stefan Dittmaier\\
        Albert-Ludwigs-Universit\"at Freiburg\\
        E-mail: \email{stefan.dittmaier@uni-freiburg.de}}

\author{\speaker{Lars Hofer}\\
        IFAE, Universitat Aut\`onoma de Barcelona\\
        E-mail: \email{lhofer@ifae.es}}

\abstract{We introduce the fortran-library \collier\ for the
        numerical evaluation of one-loop scalar and tensor integrals
        in perturbative relativistic quantum field theories. Important
        features are the implementation of dedicated methods to
        achieve numerical stability for 3- and 4-point tensor
        integrals, the support of complex masses for internal
        particles, and the possibility to choose between dimensional
        and mass regularization for infrared singularities.  \collier\ 
        supports one-loop $N$-point functions up to currently $N=6$  
        and has been tested in various NLO QCD and EW calculations.}

\FullConference{ Loops and Legs in Quantum Field Theory - LL 2014,\\
                 27 April - 2 May 2014 \\
                 Weimar, Germany }

\usepackage{amsmath,amsfonts,amssymb}
\usepackage{graphicx,epsfig,epsf}
\usepackage{times}
\usepackage{cancel}
\usepackage{axodraw}

\def\beq{\begin{equation}}
\def\eeq{\end{equation}}
\def\beqar{\begin{eqnarray}}
\def\eeqar{\end{eqnarray}}
\def\barr#1{\begin{array}{#1}}
\def\earr{\end{array}}
\def\bfi{\begin{figure}}
\def\efi{\end{figure}}
\def\btab{\begin{table}}
\def\etab{\end{table}}
\def\bce{\begin{center}}
\def\ece{\end{center}}
\def\nn{\nonumber}

\def\text{\textstyle}

\def\eps{\epsilon}
\def\veps{\varepsilon}
\def\veps{\delta}

\def\reffi#1{\mbox{Figure~\ref{#1}}}

\def\refta#1{\mbox{Table~\ref{#1}}}

\def\citere#1{\mbox{Ref.~\cite{#1}}}

\newcommand{\ri}{{\mathrm{i}}}
\newcommand{\rd}{{\mathrm{d}}}

\def\ie{i.e.,\ }

\def\ea{et al.}

\newcommand{\UV}{{\mathrm{UV}}}
\newcommand{\IR}{{\mathrm{IR}}}

\newcommand{\fin}{{\mathrm{fin}}}

\newcommand{\collier}{{\sc Collier}}
\newcommand{\coli}{{\tt Coli}}
\newcommand{\DD}{{\tt DD}}
\newcommand{\tensors}{{\tt tensors}}

\newcommand{\sst}{\scriptstyle}

\def\mathswitchr#1{\relax\ifmmode{\mathrm{#1}}\else$\mathrm{#1}$\fi}

\newcommand{\PW}{\mathswitchr W}

\newcommand{\PH}{\mathswitchr H}
\newcommand{\Pp}{\mathswitchr p}

\newcommand{\Pb}{\mathswitchr b}

\newcommand{\Pt}{\mathswitchr t}

\newcommand{\Pep}{\mathswitchr {e^+}}
\newcommand{\Pem}{\mathswitchr {e^-}}

\begin{document}

\section{Introduction}
\label{se:intro}

Next-to-leading order (NLO) predictions for processes induced by strong (QCD) and electroweak (EW) interactions are a basic ingredient
for the analysis of high-energy collider experiments. In the past
years many automatic tools based on different methods have been
developed for the calculation of QCD
corrections~\cite{QCDgen,OpenLoops}, and an NLO generator for EW
corrections has been constructed recently~\cite{Recola}. While in
unitarity-based methods~\cite{unitarity} a one-loop amplitude is
directly expressed in terms of a set of basic scalar integrals, the
traditional Feynman-diagrammatic approach as well as recently
developed recursive methods~\cite{OpenLoops,Recola,vanHameren:2009vq}
rely instead on tensor integrals. For the reduction of tensor
integrals to scalar integrals various methods have been invented and
refined over the past decades
\cite{Passarino:1978jh,reduction,Denner:2002ii,Denner:2005nn},
resulting in several libraries that are available for the calculation
of one-loop scalar and tensor integrals~\cite{looplib}. In this
article we introduce {\sc Collier}, a Complex One-Loop LIbrary in
Extended Regularisations.  Its particular strengths are the
numerically stable calculation of 3- and 4-point tensor integrals
owing to the implementation of sophisticated expansion methods for
critical phase-space regions, the support of complex masses for
internal particles, and the possibility to treat infrared
singularities either via dimensional or via mass regularisation.
Tensor integrals for 5-point and 6-point functions are reduced with
methods that do not involve inverse Gram determinants.  The library
has already been applied successfully to many complex NLO QCD and EW
calculations, among others to the processes~\cite{ee4f,processes}
$\Pep\Pem\to\PW\PW\to4\,$fermions, $\PH\to4\,$fermions,
$\Pp\Pp\to\Pt\bar\Pt\Pb\bar\Pb$, $\Pp\Pp\to\PW\PW\Pb\bar\Pb$,
$\Pp\Pp\to\Pt\bar\Pt+2{}$jets,
and $\Pp\Pp\to\ell\ell+2{}$jets. It is integrated in the NLO generators
{\sc OpenLoops}~\cite{OpenLoops} and {\sc Recola}~\cite{Recola} and
the publication of the code is in preparation~\cite{Collier}.

\section{Representation of tensor integrals}
\label{se:TIs}

A one-loop $N$-point tensor integral of rank $P$ has the general form
\beq
\label{tensorint}
T^{N,\mu_{1}\ldots\mu_{P}}(p_{1},\ldots,p_{N-1},m_{0},\ldots,m_{N-1})=
\displaystyle{\frac{(2\pi\mu)^{4-D}}{\ri\pi^{2}}\int \rd^{D}q\,
\frac{q^{\mu_{1}}\cdots q^{\mu_{P}}}
{N_0N_1\ldots N_{N-1}}}.
\eeq
The denominator factors are given by
\beq \label{D0Di}
N_{k}= (q+p_{k})^{2}-m_{k}^{2}+\ri\veps, \qquad k=0,\ldots,N-1 ,
\qquad p_0=0,
\eeq
where $p_k$ and $m_k$ are the momentum and the mass of the
particle in the corresponding loop-propagator and $\ri\veps$
$(\veps>0)$ is an infinitesimal imaginary part. While {\collier}
accepts only real values for the four-momenta $p_k$, it permits
complex values for the masses $m_k$. Thus, it can be applied to
calculations in which propagators of unstable particles are
regularised by a complex mass prescription~\cite{ee4f,cmass}. Lorentz
covariance allows to decompose a tensor integral as
\beqar
T^{N,\mu_1\ldots\mu_P} &=& 
\sum_{n=0}^{\left[\frac{P}{2}\right]} \;\;
\sum_{i_{2n+1},\ldots,i_P=1}^{N-1} \,
\{\underbrace{g \ldots g}_n  p\ldots p\}^{\mu_1\ldots\mu_P}_{i_{2n+1}\ldots i_P}
\, T^N_{\underbrace{\sst 0\ldots0}_{2n} i_{2n+1}\ldots i_{P}}
\eeqar
where $\left[{P}/{2}\right]$ is the largest integer number smaller 
or equal to ${P}/{2}$ and where the basic tensor structures are recursively defined 
according to
\beqar
\{p\ldots p\}^{\mu_1\ldots\mu_P}_{i_1\ldots i_P} &=& 
p_{i_1}^{\mu_1}\ldots p_{i_P}^{\mu_P},\label{eq:MomTen}
\nn\\[.3em]
\{\underbrace{g \ldots g}_n  p\ldots p\}^{\mu_1\ldots\mu_P}_{i_{2n+1}\ldots i_P}
&=& \frac{1}{n}
\sum_{\substack{k,l=1\\k<l}}^P \, g^{\mu_k\mu_l}
\{\underbrace{g \ldots g}_{n-1}  p\ldots p\}^{\mu_1\ldots\mu_{k-1}\mu_{k+1}\ldots\mu_{l-1}\mu_{l+1}\ldots\mu_P}_{i_{2n+1}\ldots i_P}.
\eeqar
Since the tensor $T^{N,\mu_1\ldots\mu_P}$ is totally symmetric, the Lorentz-invariant coefficients $T^N_{0\ldots0 i_{2n+1}\ldots i_{P}}$ are symmetric in $i_{2n+1},...,i_P$.

Ultraviolet- (UV-) or infrared- (IR-) singular integrals are
represented in dimensional regularisation, where $D=4-2\eps$, as
\beqar
\label{eq:normalization}
T^N&=&
T^N_{\fin}(\mu^2_\UV, \mu^2_\IR)
+  a^{\UV}\Delta_{\UV}
+ a^{\IR}_2 \left(\Delta^{(2)}_{\IR}+\Delta^{(1)}_{\IR}\ln\mu^2_\IR\right)
+ a^{\IR}_1 \Delta_{\IR}^{(1)},
\eeqar
with
\beq
\Delta_{\UV} = \frac{c(\eps_\UV)}{\eps_{\UV}}, \qquad
\Delta^{(2)}_{\IR} = \frac{c(\eps_\IR)}{\eps_{\IR}^2}, \qquad
\Delta^{(1)}_{\IR} = \frac{c(\eps_\IR)}{\eps_{\IR}}.
\label{eq:normalizationIR}
\eeq
Note that we distinguish between singularities resulting from the IR
and from the UV domain and that we absorb a term
$c(\eps)=\Gamma(1+\eps)(4\pi)^\eps$ in the constants $\Delta_{\UV}$,
$\Delta^{(2)}_{\IR}$ and $\Delta^{(1)}_{\IR}$.  {\collier} provides
numerical results for the complete integrals $T^ N$, i.e.\ for the sum
of the finite part $T^N_{\fin}(\mu^2_\UV, \mu^2_\IR)$ and the $a^{\UV}$-, $a^{\IR}_2$- and
$a_1^{\IR}$-terms. The user can assign arbitrary values to the
unphysical mass scales $\mu^2_\UV$, $\mu^2_\IR$ as well as to the
constants $\Delta_{\UV}$, $\Delta^{(2)}_{\IR}$, and
$\Delta^{(1)}_{\IR}$, which have to drop out in UV- and IR-finite
quantities. Varying these parameters allows to check numerically the
cancellation of singularities.

UV- and IR-singular integrals are by default calculated in dimensional
regularisation. Collinear singularities can also be regularised with
small masses. To this end, masses must be declared {\it small} in the
initialisation together with corresponding (not necessarily small) numerical
values. The {\it small} masses are treated as infinitesimally small in the scalar and
tensor functions, and only in mass-singular logarithms the finite
values are kept.

A general one-loop amplitude $\delta\cal{M}$ can be written in terms of tensor integrals as
\begin{equation}
   \delta {\cal{M}}\,=\,\sum_j\sum_{P_j}c^{j}_{\mu_1\cdots\mu_{P_j}}T_j^{N_j,\mu_1\cdots\mu_{P_j}}
           \,=\,\sum_j\sum_{P_j}\sum_{n=0}^{\left[\frac{P_j}{2}\right]} \;\;
\sum_{i_{2n+1},\ldots,i_{P_j}=1}^{N-1} \,
\tilde{c}^{j}_{\underbrace{\sst 0\ldots0}_{2n} i_{2n+1}\ldots i_{P_j}}
 T^{N_j}_{j,\underbrace{\sst 0\ldots0}_{2n} i_{2n+1}\ldots i_{P_j}},
\end{equation}
where $j$ runs over all appearing tensor integrals with rank $P_j$ and
$N_j$ propagators.  Traditional calculations rely on the
representation of $\delta\cal{M}$ in terms of the
$T^{N_j}_{j,i_{1}\ldots i_{P_j}}$ and perform algebraic manipulations
of the corresponding coefficients $\tilde{c}^{j}_{i_{1}\ldots
  i_{P_j}}$ in $D$ dimensions. New methods inspired by
Ref.~\cite{vanHameren:2009vq} and implemented in the automatic NLO
generators {\sc OpenLoops}~\cite{OpenLoops} and {\sc
  Recola}~\cite{Recola}, on the other hand, make use of the
representation in terms of the full tensors
$T_j^{N_j,\mu_1\cdots\mu_{P_j}}$ and perform a recursive numerical
calculation of the respective coefficients
$c^{j}_{\mu_1\cdots\mu_{P_j}}$. {\collier} can be used in either of
these approaches as it provides the Lorentz-covariant coefficients
$T^{N_j}_{j,i_{1}\ldots i_{P_j}}$ as well as the full tensors
$T_j^{N_j,\mu_1\cdots\mu_{P_j}}$.

\section{Implemented methods}
\label{se:methods}
The method used to evaluate a tensor integral depends on the number
$N$ of its propagators. For $N=1,2$, explicit numerically stable
expressions are employed~\cite{Passarino:1978jh,Denner:2005nn}.

For $N=3,4$, scalar integrals are calculated using analytical
expressions as given in \citere{scalarints}, while
tensor integrals $T^{N,P}$ of higher rank $P$ by default are
numerically reduced to integrals of lower rank $T^{N,P-1}$,
$T^{N,P-2}$ and to integrals with a lower number of propagators
$T^{N-1}$ via standard Passarino--Veltman reduction. Schematically
this can be written as
\begin{equation}
 \Delta T^{N,P}\,=\,\left[
  T^{N,P-1},T^{N,P-2},T^{N-1}\right],
  \label{eq:PV}
\end{equation}
where $[...]$ denotes a linear combination of the corresponding terms
and the determinant $\Delta=\det(Z)$ of the Gram matrix
$Z_{ij}=2p_ip_j$ has been made explicit on the left-hand side. In
certain regions of the phase-space the Gram determinant $\Delta$ can
become small, so that the numerical solution of (\ref{eq:PV}) gets
unstable. This problem reflects the ambiguity of the representation of
$T^{N,P}$ in terms of the integrals on the right-hand side which tend
to become linearly dependent in this case. Since even the scalar
integrals become dependent, this problem is intrinsic to all reduction
methods relying on the full set of basic scalar integrals, i.e.\ it
affects unitarity-based approaches as well. In the tensor reduction
method, on the other hand, spurious Gram singularities can be avoided
for delicate phase-space points by adjusting the strategy of solving
the system of linear equations obtained from the Passarino--Veltman
algorithm. Consider to this end (\ref{eq:PV}) for $P\to P+1$,
\begin{equation}
 \Delta T^{N,P+1}\,=\,\left[
  T^{N,P},T^{N,P-1},T^{N-1}\right],
  \label{eq:PV2}
\end{equation} 
in which the integral of interest, $T^{N,P}$, now appears on the
right-hand side. Neglecting in first approximation terms of order
${\cal{O}}(\Delta)$, the integrals $T^{N,P}$ can be calculated
recursively from integrals of lower rank $T^{N,P-1}$ and from
integrals with a lower number of propagators $T^{N-1}$. In this way
tensor integrals of arbitrary rank can be determined at zeroth order
in the small parameter $\Delta$. Inserting afterwards the
so-determined higher-rank tensor integral $T^{N,P+1}$ into the left-hand
side of (\ref{eq:PV2}) allows to calculate also terms of order
${\cal{O}}(\Delta)$ for $T^{N,P}$. Proceeding systematically in this
way one obtains $T^{N,P}$ as a series expansion in the parameter
$\Delta$, where higher precision in the form of ${\cal{O}}(\Delta^k)$
terms is achieved at the prize of calculating higher-rank tensor
integrals $T^{N,P+k}$.

Based on the described strategy, various expansion methods have been
suggested in Ref.~\cite{Denner:2005nn} with the respective expansion
parameter(s) depending on the region in phase space. All these methods
have been implemented in {\collier} to arbitrary order in the
expansion parameter. In order to decide which method to use for a
certain phase-space point, an a priori error estimate is performed for
the different methods considering a simplified propagation of errors
from scalar integrals and neglected higher-order terms into the tensor
integrals of highest rank. During the actual calculation of an
expansion the precision is further checked by analysing the correction
of the last iteration. In single cases where the a priori
error estimate turns out as having been too optimistic, other
expansions are tried in addition. In this way stable results are
obtained for almost all phase-space points ensuring reliable Monte
Carlo integrations.

For $N=5,6$, tensor integrals are directly reduced to integrals with
lower rank and lower $N$ following
Refs.~\cite{Denner:2002ii,Denner:2005nn}, \ie without using inverse
Gram determinants. The methods summarised there can be extended to the
case of $N\ge 7$ in a straightforward way.

While the methods described so far are formulated in the literature in
terms of the Lorentz-invariant coefficients $T^N_{i_1\ldots i_P}$, a
new generation of NLO generators, such as {\sc OpenLoops} and 
\linebreak
{\sc
  Recola}, needs the elements of the full tensors
$T^{N,\mu_1\cdots\mu_{P}}$. To this end, an efficient algorithm has
been implemented in {\collier} to construct the tensors
$T^{N,\mu_1\cdots\mu_{P}}$ from the coefficients $T^N_{i_1\ldots
  i_P}$. It performs a recursive calculation of those tensor
structures in (\ref{eq:MomTen}) that are built exclusively from momentum
vectors. Non-vanishing elements of other tensor structures involving
metric tensors are then obtained by adding pairwise equal
Lorentz indices, and their value differs from the corresponding value
of the pure momentum tensor only by a combinatorial factor and a
potential minus-sign induced by the metric tensors. The relevant
combinatorial factors are calculated and tabulated during the
initialisation of {\collier}.
\begin{table}[t]
\let\Green\relax
\let\Red\relax
\let\Blue\relax
\small
  $$\begin{array}{c@{\quad}|c@{\quad}c@{\quad}c@{\quad}c@{\quad}c@{\quad}c@{\quad}c}
             &  \Blue{P=0}  & \Blue{P=1}  & \Blue{P=2} & \Blue{P=3} & \Blue{P=4} & \Blue{P=5} & \Blue{P=6}\\[.3ex]\hline
\Green{N=3}          &  1    &  2   &   4  &  6  &   9 &  12  &   16\\[.3ex]
\Green{N=4}          &  1    &  3   &   7  & 13  &  22 &  34  &   50\\[.3ex]
\Green{N=5}          &  1    &  4   &  11  & 24  &  46 &  80  &  130\\[.3ex]
\Green{N=6}          &  1    &  5   &  16  & 40  &  86 & 166  &  296\\[.3ex]
\Green{N=7}          &  1    &  6   &  22  & 62  & 148 & 314  &  610\\[.3ex]\hline
\Red{\rm{tensor}}    &  1    &  4   &  10  & 20  &  35 & 56  &  84 
\end{array}$$\label{tab:numbers}
\caption{Number of invariant coefficients $T^N_{i_1\ldots i_P}$ for
  $N=3,...,7$ and rank $P=0,...,6$ (rows 2-6) and number of tensor
  elements $T^{N,\mu_1\cdots\mu_{P}}$ for rank $P=0,...,6$ (last
  row)}.  
\end{table}

The numbers of invariant coefficients $T^N_{i_1\ldots i_P}$ and tensor
elements $T^{N,\mu_1\cdots\mu_{P_j}}$ are compared in
\refta{tab:numbers}. For $N\le 4$ the number of invariant coefficients
is smaller than the number of tensor elements, and this fact
constitutes a basic precondition of the Passarino--Veltman reduction
method. For $N\ge 5$, on the other hand, there are less tensor
elements than coefficients and the reduction method for $N\ge 6$
presented in (7.7) of Ref.~\cite{Denner:2005nn} has been actually
derived in terms of full tensors. Its translation to tensor
coefficients requires an additional symmetrisation and the resulting
coefficients are not unique because of the overdefined number of
tensor structures. Therefore for the calculation of the tensors
$T^{N,\mu_1\cdots\mu_{P}}$ the reduction for $N\ge 6$ has been
implemented in {\collier} also directly at the tensor level without
resorting to a covariant decomposition.

\section{Structure of the library}
\label{se:structure}
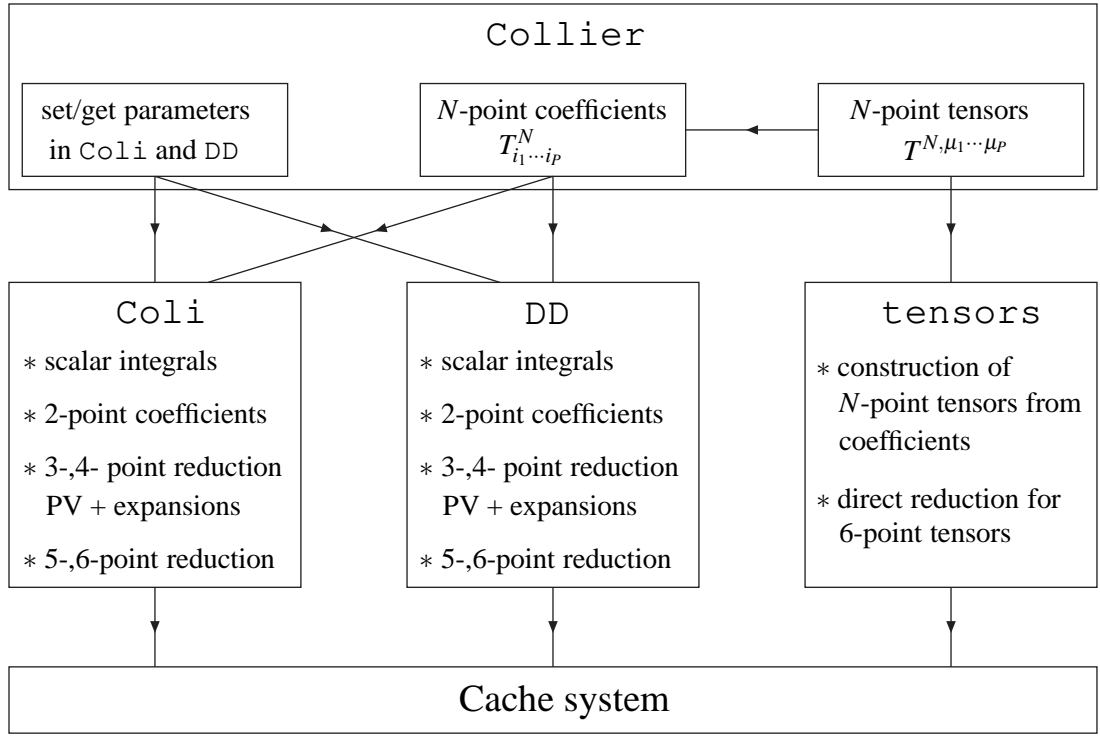
\begin{figure}[t]
  \begin{picture}(200,280)(-80,-30)
    \SetColor{Black}

    \BBox(-70,180)(340,250)

    \BBox(-65,185)(35,220)
    \BBox(85,185)(185,220)
    \BBox(235,185)(335,220)

    \BBox(-70,30)(40,145)
    \BBox(80,30)(190,145)
    \BBox(230,30)(340,145)
    \BBox(-70,-25)(340,0)

    \ArrowLine(-15,30)(-15,0)
    \ArrowLine(135,30)(135,0)
    \ArrowLine(285,30)(285,0)
    \ArrowLine(285,185)(285,145)
    \ArrowLine(-15,185)(-15,145)
    \ArrowLine(-15,185)(115,145)
    \ArrowLine(135,185)(135,145)
    \ArrowLine(135,185)(5,145)
    \ArrowLine(235,202.5)(185,202.5)

    \Text(110,235)[lb]{\Large{\tt Collier}}

    \Text(-30,130)[lb]{\Large{\coli}}
    \Text(125,130)[lb]{\Large{\DD}}
    \Text(260,130)[lb]{\Large{\tensors}}
    \Text(100,-20)[lb]{\Large{Cache system}}

    \Text(-65,110)[lb]{\normalsize{$*$ scalar integrals}}
    \Text(-65,90)[lb]{\normalsize{$*$ 2-point coefficients}}
    \Text(-65,70)[lb]{\normalsize{$*$ 3-,4- point reduction}}
    \Text(-65,55)[lb]{\normalsize{\;\; PV + expansions}}
    \Text(-65,35)[lb]{\normalsize{$*$ 5-,6-point reduction}}

    \Text(85,110)[lb]{\normalsize{$*$ scalar integrals}}
    \Text(85,90)[lb]{\normalsize{$*$ 2-point coefficients}}
    \Text(85,70)[lb]{\normalsize{$*$ 3-,4- point reduction}}
    \Text(85,55)[lb]{\normalsize{\;\; PV + expansions}}
    \Text(85,35)[lb]{\normalsize{$*$ 5-,6-point reduction}}

    \Text(235,110)[lb]{\normalsize{$*$ construction of}}
    \Text(235,94)[lb]{\normalsize{\;\; $N$-point tensors from}}
    \Text(235,82)[lb]{\normalsize{\;\; coefficients}}
    \Text(235,60)[lb]{\normalsize{$*$ direct reduction for}}
    \Text(235,45)[lb]{\normalsize{\;\; 6-point tensors}}

    \Text(-58,205)[lb]{\normalsize{set/get parameters}}
    \Text(-55,192)[lb]{\normalsize{in {\coli} and {\DD}}}

    \Text(92,205)[lb]{\normalsize{$N$-point coefficients}}
    \Text(115,190)[lb]{\normalsize{$T^N_{i_1\cdots i_P}$}}

    \Text(248,205)[lb]{\normalsize{$N$-point tensors}}
    \Text(268,192)[lb]{\normalsize{$T^{N,\mu_1\cdots \mu_P}$}}
  \end{picture}\label{fig:structure}
\caption{Structure of the library {\collier}.}
\end{figure}

The structure of the library {\collier} is illustrated schematically
in \reffi{fig:structure}. The core of the library is formed by the
building blocks {\coli} and {\DD}. They constitute two independent
implementations of the scalar integrals $T^N_0$ and the
Lorentz-invariant coefficients $T^N_{i_1\ldots i_P}$ employing the
methods described in the previous section. The module {\tensors}
provides routines for the construction of the tensors
$T^{N,\mu_1\ldots \mu_P}$ from the coefficients $T^N_{i_1\ldots i_P}$
as well as for a direct reduction of 6-point integrals at the tensor
level. The user interacts with the basic routines of {\coli}, {\DD}
and {\tt tensors} via the global interface of {\collier}. It provides
routines to set or extract numerical values of the parameters in
{\coli} and {\DD} as well as routines to call the calculation of
tensor coefficients $T^N_{i_1\ldots i_P}$ or tensor elements
$T^{N,\mu_1\ldots \mu_P}$. The user can choose whether the {\coli}- or the
{\DD}-branch shall be used for the calculation of the integrals. It is
also possible to calculate each integral with both branches 
for the purpose of comparison.

In the evaluation of a one-loop matrix element the same tensor
integral is called various times: On the one hand, a single user call
of an $N$-point integral leads to recursive internal calls of lower
$N^{\prime}$-point integrals and for $N^{\prime}\le N-2$ the same
integral is reached through more than one path in the reduction tree.
On the other hand, different user calls and their reductions typically
involve identical tensor integrals. In order to avoid multiple
calculations of the same integral the sublibraries of {\collier} are
linked to a global cache system which works as follows: A parameter
$N_{\rm ext}$ numerates external integral calls, while for the
book-keeping of internal calls a binary identifier $id$ is propagated
during the reduction. A pointer is assigned to each index pair
$(N_{\rm ext},id)$. During the evaluation of the first phase-space
points the arguments of the corresponding function calls are compared
and pairs $(N_{\rm ext},id)$ with identical arguments are pointed to
the same address in the cache. For later phase-space points the result
of the first call of an integral is written to the cache and read out
in subsequent calls pointing to the same address.

\section{Conclusions}
\label{se:conclusions}
We have introduced the fortran-based Complex One-Loop LIbrary in
Extended Regularizations {\collier}. It provides the complete set of
basic scalar integrals as well as tensor integrals of arbitrary rank
for up to $N=6$ external particles (an implementation for $N\ge 7$ is
in progress).

In order to ensure numerical stability the expansion methods for 3-
and 4-point integrals of Ref.~\cite{Denner:2005nn} have been
implemented to arbitrary order in the corresponding expansion
parameter. UV singularities are regularised dimensionally, IR
singularities integrals can be regularised dimensionally or
alternatively by introducing small masses. Complex values are
supported for the masses of internal particles in loop propagators,
permitting thus the application of {\collier} to processes involving
unstable particles. As output the user obtains either the coefficients
$T^N_{i_1\ldots i_P}$ of the covariant decomposition of the respective
tensor integral or the elements of the tensor $T^{N,\mu_1\ldots
  \mu_P}$ themselves. A recalculation of identical integrals is
avoided by an efficient built-in cache system. The fundamental
building blocks of the library are provided in two implementations
that allow for an independent calculation of each integral and for
direct numerical cross-checks.

{\collier} has already been successfully applied to a large number of calculations of QCD and EW corrections and is integrated in the NLO generators {\sc OpenLoops} and {\sc Recola}. Publication of the code facilitating its use by other generators and other groups is in preparation.  

\subsection*{Acknowledgements}
This work was supported in part by the Deutsche Forschungsgemeinschaft
(DFG) under reference number DE~623/2-1.  The work of L.H. was
supported by the grant FPA2011-25948.

\end{document}